\documentclass[12pt]{iopart}
\usepackage{iopams}
\usepackage{graphicx}

\begin{document}

\title[Photon production from non-equilibrium QGP in heavy ion collisions]
{Photon production from non-equilibrium QGP\\ in heavy ion collisions}

\author{ F Gelis\dag,  \underline{H Niemi}\ddag, P V Ruuskanen\ddag\ and S S  R\"as\"anen\ddag}

\address{\dag\ Service de Physique Th\'eorique,
CEA/DSM/Saclay, \\
F-91191 Gif-sur-Yvette cedex, France}

\address{\ddag\ Department of Physics,  University of Jyv\"askyl\"a, \\
        P.O. Box 35, FIN-40014 University of Jyv\"askyl\"a, Finland}

\begin{abstract}
We present a calculation of thermal photon production i.e. photons from
secondary interactions among particles produced in heavy ion collisions
at collider energies.  This is done within the framework of
hydrodynamics.  We take into account the lack of chemical equilibrium in
QGP.  It turns out that main effects from chemical
non-equilibrium
composition of QGP, reduction of particle number and increase in
temperature, nearly cancel in photon spectrum.
 \end{abstract}




\section{Introduction}

Thermal emission of real photons is very sensitive to the initial
temperature of emitting matter and their  mean
free path in the produced matter is so long that they escape
from the fireball without further interactions.  This makes thermal
photons an excellent probe of the early stage of the collision.
Unfortunately, even in best situations,
thermal photons are only a small contribution to the total photon
spectrum.  At low transverse momentum $k_T$, the main contribution is
from the electromagnetic decays of hadrons, mostly from $\pi^0$, and at
high $k_T$ from prompt photons i.e. photons from primary interactions of
the partons of the colliding nuclei.  At collider energies, however, we
have a possible window at intermediate $k_T \simeq 3-7$ GeV to detect
the thermal part of the photon spectrum.  Also a new contribution to the
photon spectrum has been proposed recently~\cite{Fries, Gale}.  As high energy
jets propagate through QGP they will induce emission of real photons.
Photons from this process will mix with the thermal
spectrum increasing the overall signal from QGP.

One uncertainty
in thermal emission is the chemical composition of the QGP.
We present here a calculation of the thermal part of the
photon spectrum within the framework of hydrodynamics~\cite{frhydro}.
Supplemented with photon emission rates in thermal matter, hydrodynamics
provides a closed framework to study the thermal emission in a
heavy-ion collision.
Within this framework it is relatively easy to take the
chemical composition of the QGP into account.  It turns out that the
effects from chemical non-equilibrium, the decrease
of quark number relative to gluons and the increase in temperature,
nearly cancel in the thermal photon contribution to the spectrum.

\section{Theoretical framework}

Our framework to study the thermal photon emission is the following:
The initial energy and particle densities are obtained from the pQCD +
saturation model~\cite{EKRT} which also gives the production time of the
initial parton matter.  Assuming thermal equilibrium immediately after
production, the whole expansion of the matter can be treated by using
hydrodynamics.  This model, with the extra assumptions of longitudinal
boost invariance and that the matter behaves as an ideal fluid, is in
good agreement with the measured hadron spectra at RHIC~\cite{ENRR}.
Since we study the effects of particle composition on the photon
emission, we relax the assumption of chemical equilibrium. We also
neglect the small net baryon number.

Chemical composition of the matter can be controlled by introducing
chemical potentials which we approximate  by using
multiplicative fugacities
for both quarks and gluons.  The time evolution of fugacities is
determined
by rate equations and following Biro et al.~\cite{Biro}, processes
$q\bar{q}
\leftrightarrow gg \,\,\,{\rm and} \,\,\, gg \leftrightarrow ggg $
are included. Rate equations can be solved together with the
hydrodynamic equations~\cite{Biro, Elliott, Srivastava}, once the
equation of state is specified.  This model gives a complete space-time
evolution of energy density and transverse velocity as well as
temperature, fugacities and other thermodynamic quantities.
Once the space-time evolution is known from hydrodynamics
and the photon emission rate as a function of temperature
and fugacities is specified, we can integrate the rate over the
space-time history of the collision to obtain the spectrum of thermal
photons.  Photon emission rate in the hot hadron gas is calculated and
parametrized in refs.~\cite{Kapusta, Xiong, Nadeau}.  In QGP in chemical equilibrium the
emission rate, complete to leading order in $\alpha_{em}$ and
$\alpha_s$, is calculated in~\cite{AMY2}.
Following the method of ref.~\cite{AGMZ}, this calculation has been
extended to the non-equilibrium case in ref.~\cite{fugacity}.

\section{Equation of state}
The equation of state (EoS) is specified by combining the hadron gas equation of
state and the QGP bag model EoS by Maxwell construction. The hadron gas is an
ideal gas of all hadronic states up to a mass $1.3$
GeV and QGP is treated as an ideal gas of gluons and three flavors of
massless quarks.  In the approximation of multiplicative fugacities
$\lambda_i$ for quarks and gluons the particle densities in the QGP are
given by
\begin{equation}
n_i(T, \lambda_i) = \lambda_i \hat{n}_i(T) = \lambda_i a_i T^3,\,\, i = q, g
\label{eq:nd}
\end{equation}
where $\hat{n}_i$ is the particle density in chemical equilibrium.
Within the same approximation, pressure is given by
\begin{equation}
p = \lambda_q \hat{p}_q + \lambda_g \hat{p}_g - B, \label{eq:pr}
\end{equation}
where $B$ is bag constant.  The connection between the critical
temperature $T_c$ and the bag constant $B$ is given by Gibbs phase
co-existence condition
\begin{equation}
p_{HG}(T_c) = p_{QGP, th}(T_c, \lambda_q, \lambda_g) - B(\lambda_q, \lambda_g),
\end{equation}
where $p_{QGP, th}$ is the thermal part of the pressure in
QGP~(\ref{eq:pr}).

The use of fugacities for describing the gluon and quark densities out
of chemical
equilibrium leads to an ambiguity on how to choose the
bag constant
and the critical temperature (cf.~eq.~(3)).  In the full kinetic and
chemical equilibrium with fugacities equal to one the situation is clear
since the temperature is the only independent variable (neglecting net
baryon number) in both phases.  With QGP out of chemical equilibrium, we
do not know quantities in the hadron gas which would be uniquely defined
from the quark and gluon fugacities at phase transition.  Here we fix
the ambiguity in matching by simply taking $T_c$ to be a constant,
independent of fugacities.  In other words, we assume that the QGP
always hadronizes at the same temperature into the same state of hadron
gas.  In our calculations $T_c = 167$ MeV both in and out-of chemical
equilibrium.
Because the photon emission from QGP is dominated by the early, high
density stage of the collision our results are not sensitive to the
details of the phase transition.

\section{Results}

The time evolution of temperature and fugacities in the center of the
fireball are shown in the Fig.~1.
As can be seen from the top figure the matter is close to chemical
equilibrium at the end of the QGP phase.  This justifies further
our choice of fixed value for $T_c$.  In the lower figure we can see the
time evolution of temperature compared with the full chemical
equilibrium
case.  Same initial energy density is used both in equilibrium and
in out-of-equilibrium case.

These two figures show the effect of lack of
chemical equilibrium.  When the energy density is kept constant, reduction of
the particle number below its equilibrium value leads to an increase of
temperature.  Under these conditions there are less particles but they
have higher average energy.  These deviations from the equilibrium
values have opposite effects on the photon emission rates and nearly
cancel each other.  This can be seen from Fig.~2, where we plot thermal
photon spectrum for both cases.  Only at high transverse momenta,
$k_T\gtrsim 10$ GeV, the increase in temperature is becoming more
important and the equilibrium spectrum starts to fall faster than the
non-equilibrium one.  However the experimentally accessible region is
around $k_T \sim 5$ GeV, where the spectrum is practically independent
of the chemical composition of QGP.

\begin{figure}
\hspace{-2.6cm}
\begin{minipage}[b]{0.62\linewidth}
\centering
\hspace*{+2.2cm}
\vspace{-0.15cm}
    \includegraphics[width=7.0cm]{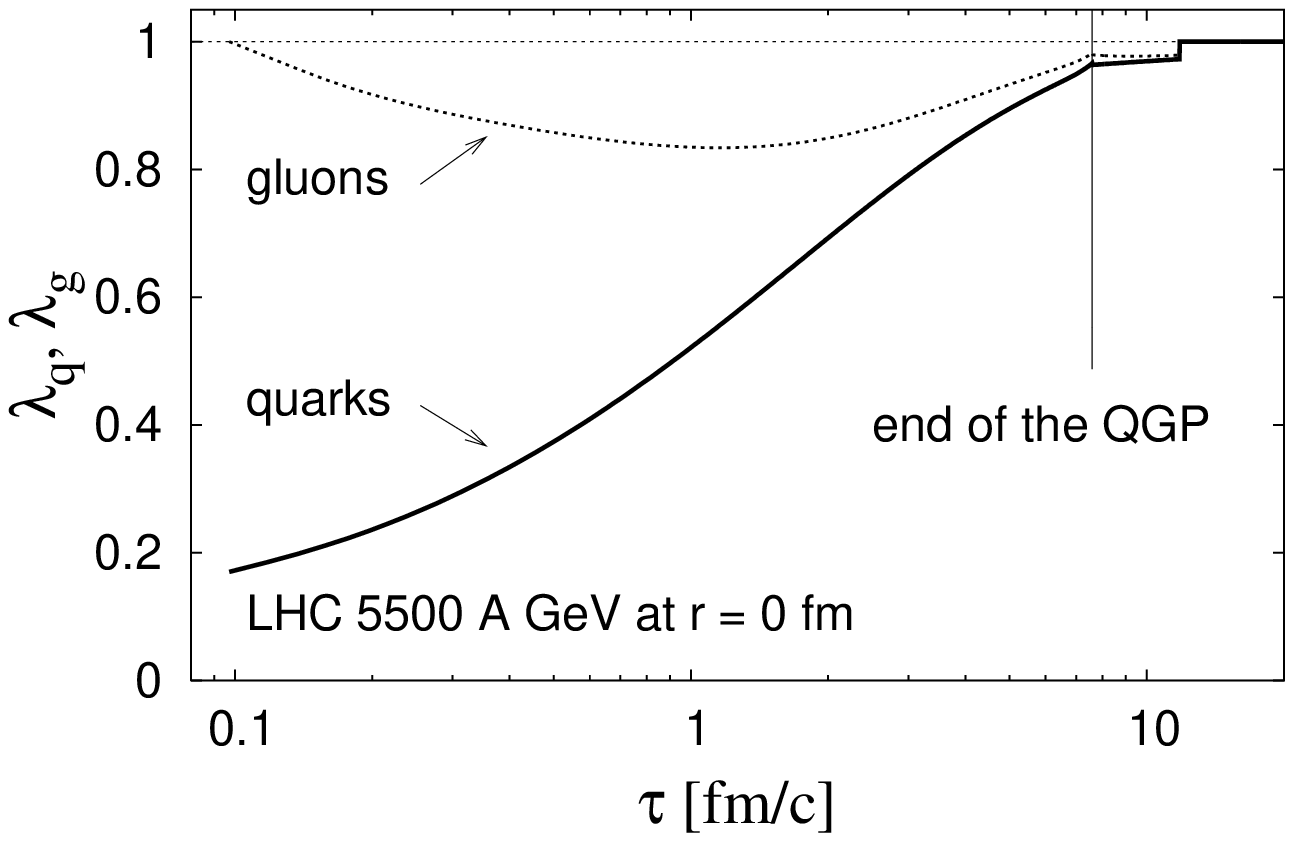}
\hspace*{+2.2cm}
    \includegraphics[width=7.0cm]{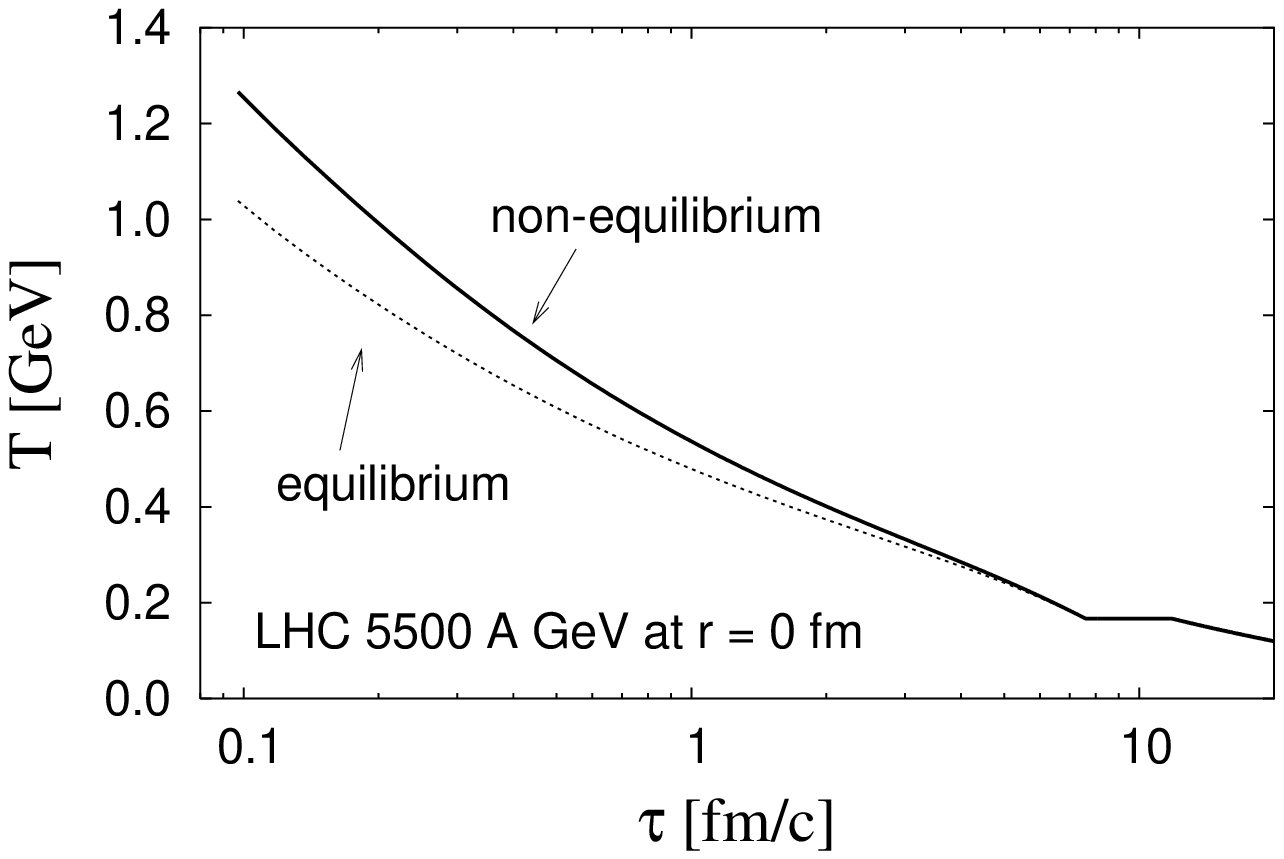}
\caption{Time evolution of fugacities and temperature in the
middle of the fireball  in the LHC lead-lead collision.}
\end{minipage}
\hspace*{-1.7cm}
\begin{minipage}[b]{0.62\linewidth}
\centering
\hspace*{-0.6cm}
\vspace*{-3.25cm}
\includegraphics[width=9.8cm,bb= -1 -1 500 500]{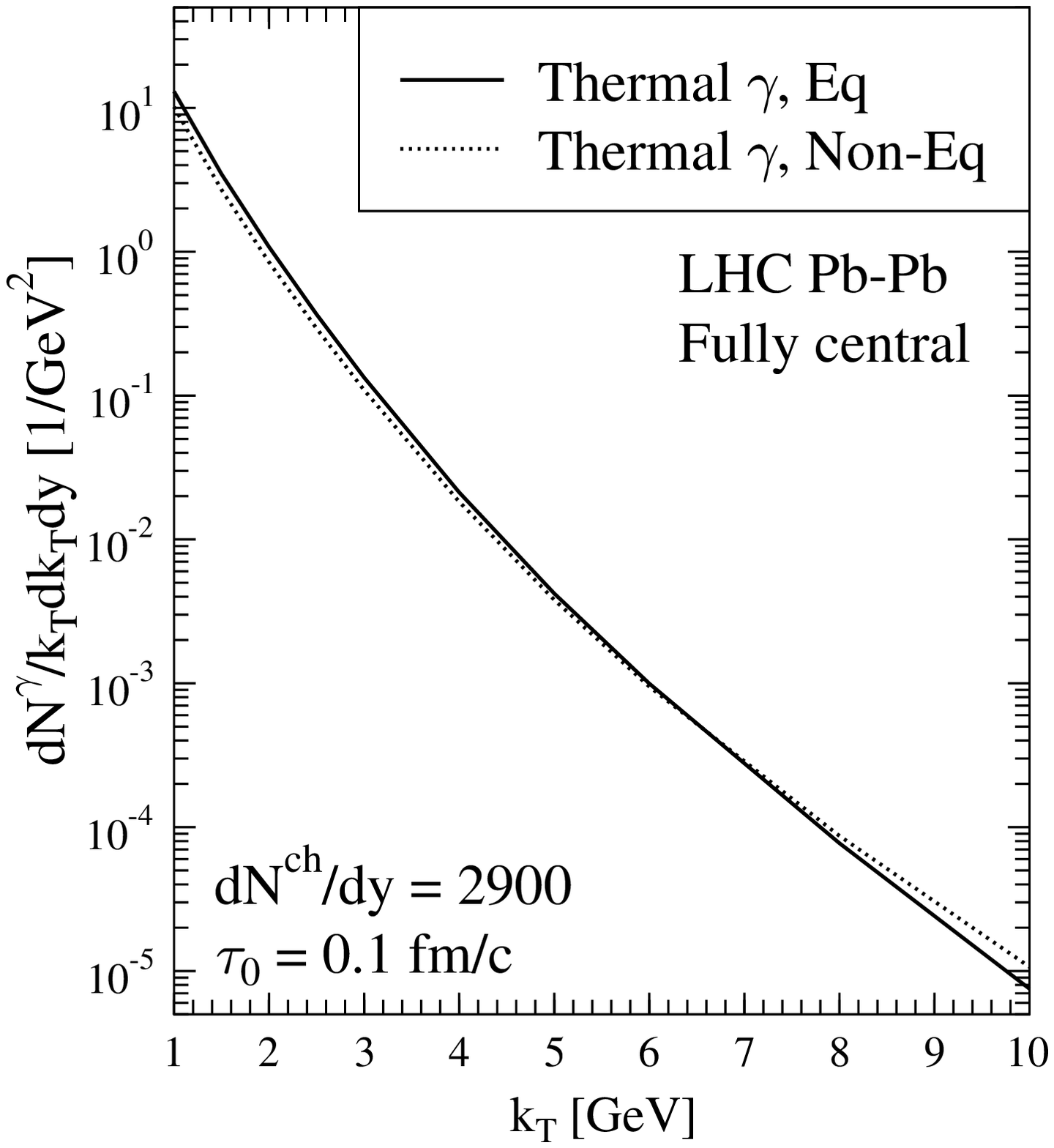}
\caption{Thermal photon spectrum for the LHC
lead-lead collision both in and out-of chemical equilibrium.}
\end{minipage}
\end{figure}

\section{Discussion}

We have calculated the spectrum of thermal photons, emitted during the
expansion stage of heavy ion collision, in the case when the QGP is not
in chemical equilibrium.  The space-time evolution of the heavy ion
collision is modeled by hydrodynamics together with rate equations for
the time evolution of chemical composition of QGP.  Thermal photon
spectrum is then obtained by integrating the temperature and in the QGP
the fugacity dependent photon emission rates over this space-time
history.
We have demonstrated that the lack of chemical equilibrium does not
change much the photon spectrum as compared to the chemical equilibrium
case.  The reason for this is that while there are less particles in the
non-equilibrium QGP, the temperature becomes higher.  These have
opposite effects on photon emission and nearly cancel in the region
where the possibilities to detect experimentally the thermal part of the
spectrum are the best.  Thus in heavy ion collisions thermal photons are
not measuring initial temperature alone, but the combination of
temperature and particle densities.

\end{document}